\documentclass{PoS}
\usepackage{url}
\usepackage{amsmath}
\usepackage{mathrsfs}
\usepackage{multirow}
\usepackage{multicol}
\usepackage{color}
\usepackage{xfrac} 
\usepackage{comment}

\usepackage{graphicx}
\usepackage{xcolor}
\usepackage{blindtext}
\usepackage{wrapfig}

\title{A Three-dimensional Reconstruction of Cosmic Ray Events in IceCube}

\ShortTitle{3-D reconstruction of cosmic ray events in IceCube}

\author{
The IceCube Collaboration\footnote{For collaboration list, see PoS(ICRC2019) 1177.}\\
{\itshape \href{http://icecube.wisc.edu/collaboration/authors/icrc19_icecube}{http://icecube.wisc.edu/collaboration/authors/icrc19\_icecube}}\\
E-mail: \email{Xinhua.Bai@sdsmt.edu}
}

\abstract{
The IceCube Neutrino Observatory at the geographic South Pole consists of two components, 
a km$^2$ surface array IceTop and a km$^3$ in-ice array between 1.5 and 2.5 km below the surface. 
Cosmic ray events with primary energy above a few tens of TeV may trigger both the IceTop and in-ice array 
and leave a three-dimensional footprint of the electromagnetic and muonic components in the extensive air 
shower. A new reconstruction based on the minimization of a unified likelihood function involving  
quantities measured by both IceTop and in-ice detectors was developed. This report describes the new 
reconstruction algorithm and summarizes its performance tested with Monte Carlo events under two 
different containment conditions. The advantages of the new reconstruction are discussed in comparison 
with reconstructions that use IceTop or in-ice data separately. Some possible improvements are also 
summarized.

\vspace{4mm}
{\bfseries Corresponding authors:} 
Xinhua Bai$^{1}$, \speaker{Emily Dvorak}$^{1}$, Javier Gonzalez$^{2}$, Dennis Soldin$^{2}$
\\
{$^{1}$ \itshape Physics Department, South Dakota School of Mines \& Technology, Rapid City, SD 57701}\\
{$^{2}$ \itshape Department of Physics \& Astronomy, University of Delaware, Newark, DE 19716}\\
}

\FullConference{36th International Cosmic Ray Conference -ICRC2019-\\
		July 24th - August 1st, 2019\\
		Madison, WI, U.S.A.}

\begin{document}
\vspace{-0.1pc} 
\section{Introduction}\label{sec_intro} 
\vspace{-0.1pc} 
The IceCube Neutrino Observatory~\cite{IceCube} at the geographic South Pole plays a crucial role in multi-messenger 
astronomy by observing high energy neutrinos from astrophysical or cosmological origins. With its 
one km$^2$ surface array IceTop and a km$^3$ in-ice array between 1.5 and 2.5 km deep in the 
glacial ice, IceCube is also a powerful detector for the study of cosmic rays (CRs), atmospheric muons, 
atmospheric neutrinos, etc. A series of science results on these topics have been obtained 
from IceCube data~\cite{IceCube_ICRCa, IceCube_ICRCb}. 

A prominent feature in astrophysical or cosmological phenomena is that the event rate often decreases 
dramatically as the energy increases. A typical example is the CR spectrum. Besides a couple of 
structures, the inverse power law index of the differential flux of CRs is between 2.7 and 3.0 over 
more than 10 orders of magnitude of energy. The study of high energy phenomena requires large 
detectors to increase the number of events. Also beneficial is to develop novel techniques that 
make maximal use of available data. A new three-dimensional (3-D) reconstruction is developed 
for the studies of CRs in IceCube. It not only largely increases 
the number of events for physics study at high energies but also provides new parameters that 
may improve the resolution of the measurement of CR primary energy and composition. The 
major software components and some updated formulas of the reconstruction are summarized in 
Section~\ref{sec_form}. Section~\ref{sec_perf} describes the performance test of the new 
reconstruction, followed by an outlook for further improvements. 

\vspace{-0.1pc} 
\section{Major software components and key formulas}\label{sec_form}
\vspace{-0.1pc} 
The 3-D reconstruction takes the size and time of signals observed by both IceTop and the in-ice array and feeds 
them into a likelihood (LH) maximization process to reconstruct the extensive air shower (EAS) profile 
and energy losses of high energy muons in the in-ice array. The data/information flow and the key software 
components are shown in Figure~\ref{fig_FlowChart}. Major service modules and operations are: 
\begin{figure}[ht]
\vspace{-0.0pc} 
\begin{center}
\fcolorbox{black}{white}{\includegraphics[width=0.94\linewidth, angle=0]{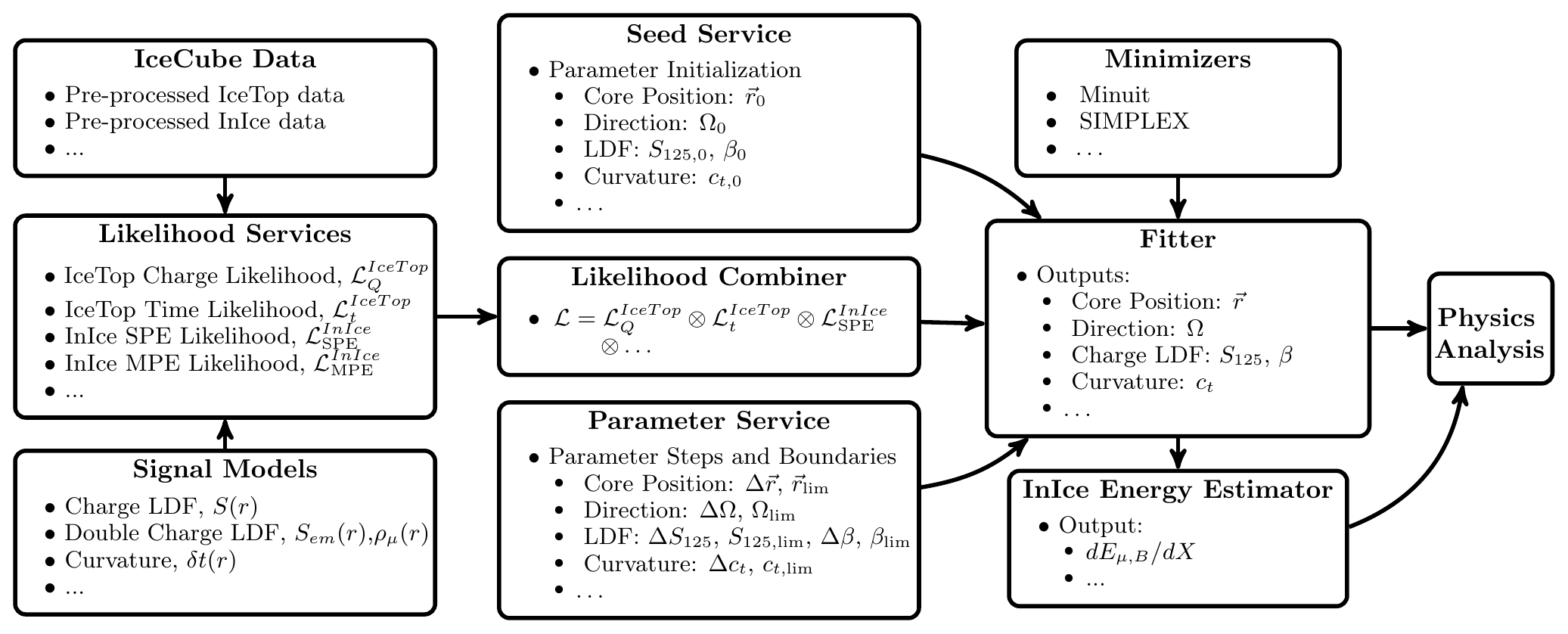}} 
\vspace{0.0pc} 
\caption{\label{fig_FlowChart} The key software components of the 3-D reconstruction and the intrinsic 
data/information flow through them. The final reconstructed EAS quantities (from the "Fitter") and muon 
energy losses (from the "InIce Energy Estimator") are used for physics analysis. 
See details of some of the indicated variables and functions in the text.} 
\end{center} 
\vspace{-1.5pc} 
\end{figure}

\noindent $\bullet ~$~\underline{Likelihood Services}: 
Provide LH descriptions for an EAS signal hypothesis in IceTop~\cite{IceTop_NIM_2013}, 
as well as for the muon signals observed in the in-ice array~\cite{MuonTrackReco_NIM_2004}.  

\vspace{0.1cm}
\noindent $\bullet ~$~\underline{Signal Models}: 
Define EAS profile by describing the lateral distribution function (LDF) $S(r)$, the 
shower curvature $\delta t(r)$, etc. for particle signals in the IceTop tanks. These models provide 
the hit probabilities in order to construct corresponding likelihood functions (LHF).

\vspace{0.1cm}
\noindent $\bullet ~$~\underline{Likelihood Combiner}: Combines the likelihoods which are used 
for the minimization procedure. It has no constraint on the number of likelihoods so that additional 
observables or new detector components can be easily implemented into the new reconstruction. 

\vspace{0.1cm}
\noindent $\bullet ~$~\underline{Seed/Parameter Service}: Initializes the parameters, defines 
the boundaries and initial step sizes for the minimization. 

\vspace{0.1cm}
\noindent $\bullet ~$~\underline{Minimizer/Fitter}: Combines all information from the services 
and conducts the minimization for the combined LH using common minimization algorithms, 
such as Minuit or Simplex~\cite{Minuit_SIMPLEX}.

\vspace{0.1cm}
\noindent $\bullet ~$~\underline{InIce Energy Estimators}: Reconstructs muon or muon bundle 
energy losses at different depths by using the reconstructed muon track information from the 
"Minimizer/Fitter" with the existing IceCube reconstruction "Millipede"~\cite{Millipede, EnerRecoMeth_JINST_2014}. 

\vspace{0.2cm}

The new framework not only has the freedom to include multiple terms in the combined LHF but 
also allows changing the functions in the LHFs for the maximum flexibility. 
In this work the combined LHF has three terms, i.e. the charge term ($\mathcal{L}^{IceTop}_{Q}$) and 
time term ($\mathcal{L}_{t}^{IceTop}$) for EAS signals in the IceTop tanks, and an InIce term 
(ex. $\mathcal{L}_{SPE}^{InIce}$, etc.) for the muon bundle footprint measured by the in-ice array. 
The new reconstruction utilizes the following LDF $S(r)$, shower front curvature function $\delta t(r)$, 
and time fluctuation $\sigma_{t_i}$: 
\begin{align} 
\label{eq_LDF}
& \text{LDF:} ~S(r)=S_{ref}\cdot \left ( \frac{r}{r_{ref}} \right )^{-\beta-0.30264 \cdot log_{10}(r/r_{ref})} \\ 
\label{eq_Curvature}
& \text{Curvature:} ~\delta t(r) = c_t r^{2} + 19.41 \cdot \left ( 1- e^{- \sfrac{r^{2}}{2\cdot (118.1))^{2}}} \right ) \text{, $c_t$ free parameter}    \\
\label{eq_Fluctuation}
& \text{Fluctuation at the $i^{th}$ station:} ~\sigma_{t_i} = C_1 \cdot \frac{\sqrt{\sum_{j=1}^{2} \left ( t_{ij}-\frac{t_{i1}+t_{i2}}{2} \right )^{2}}}{\left (\sum_{j=1}^{2}Q_{ij} \right )^{1.0}}+C_2  
\end{align} 
The definition of parameters and coefficients in Eq.~(\ref{eq_LDF}) and (\ref{eq_Curvature}) are the same with 
these used in the standard IceTop-alone reconstruction~\cite{IceTop_NIM_2013} except that the curvature parameter 
$c_t$ in Eq.~(\ref{eq_Curvature}) is a free parameter in the 3-D reconstruction instead of a fixed value. This is 
to take into account the fact that the shape of shower front curvature depends on primary energy and zenith angle. 
The new per-event based time fluctuation $\sigma_{t_i}$ in Eq.~(\ref{eq_Fluctuation}) 
is introduced so that it can be consistent with using a flexible curvature that varies for each individual EAS in the 
likelihood maximization. $t_{ij}$ and $Q_{ij}$ (with $j=1$ or $2$) are the signal time (in $ns$) and size (in $VEM$, 
vertical equivalent muon) at two IceTop tanks of the $i^{th}$ IceTop station. The values of $C_1$ 
and $C_2$ derived from a MC study are 4.0~$VEM$ and 1.22~$ns$. 
The InIce LHF is based on the probabilities of photon arrival times at DOMs, the 
signal sizes and the topology of triggered DOMs. 
The new 3-D reconstruction allows using the probability density function of single-photo-electron 
($\mathcal{L}_{SPE}^{InIce}$), or multi-photo-electrons ($\mathcal{L}_{MPE}^{InIce}$), or of other 
types. See details of the InIce LHF options in ~\cite{MuonTrackReco_NIM_2004, EnerRecoMeth_JINST_2014}. 

\vspace{-0.1pc} 
\section{Performance}\label{sec_perf} 
\vspace{-0.1pc} 
The 3-D reconstruction was tested using the IceCube MC events of the 2012 detector configuration.  
The simulated dataset is divided in two groups of events by using their MC true geometry:
events with core contained in IceTop 
{\it and} high energy muons contained in the in-ice array 
(to be referred to as IT-Contained), 
and events with core landing on the edge or outside of the IceTop 
array {\it and} high energy muon track contained in the in-ice array (IT-Uncontained). 
Although the main motivation of developing the 3-D reconstruction is to make use of IT-Uncontained 
events for physics analysis, testing it with IT-Contained events gives us the most convenience in the 
development since we have much better understanding of them through analyses 
previously done. 

The MC events use EAS generated by CORSIKA~\cite{corsika} (with Sibyll2.1~\cite{Sibyll2_1} 
as high energy interaction model), which 
include both proton and iron primaries in a zenith range between 0\textdegree ~and 
65\textdegree ~with an $E^{-1}$ energy spectrum from $10^5$ to $10^{9.5}$~GeV. 
In the production of MC events, these CORSIKA showers were evenly distributed in four circular 
areas with four different radii $R$ from the center of the IceTop array, i.e. 
$R=$1100~m, 1700~m, 2600~m and 2900~m for CRs in four primary energy bins 
$E_{pri.} \in 10^5 - 10^7$~GeV, 
$10^7 - 10^8$~GeV, 
$10^8 - 10^9$~GeV, and 
$10^9 - 10^{9.5}$~GeV.  

We also adopted quality cuts (QC) used in IceTop and InIce coincidence 
analysis~\cite{IceTop_SpectrumComposition2019} to select quality IT-Contained events. 
The standard InIce QC {\it and} five triggered IceTop stations with a station density $\geq 0.2$ 
~are used to select quality IT-Uncontained events for the 3-D reconstruction. Events are 
weighted to a $E^{-2.7}$ spectrum for all plots unless noted otherwise. Error bars on all 
plots represent standard deviation. 

\begin{wrapfigure}{r}{0.5\textwidth}
\begin{center}
\vspace{-2.0pc}
\includegraphics[width=0.48\textwidth]{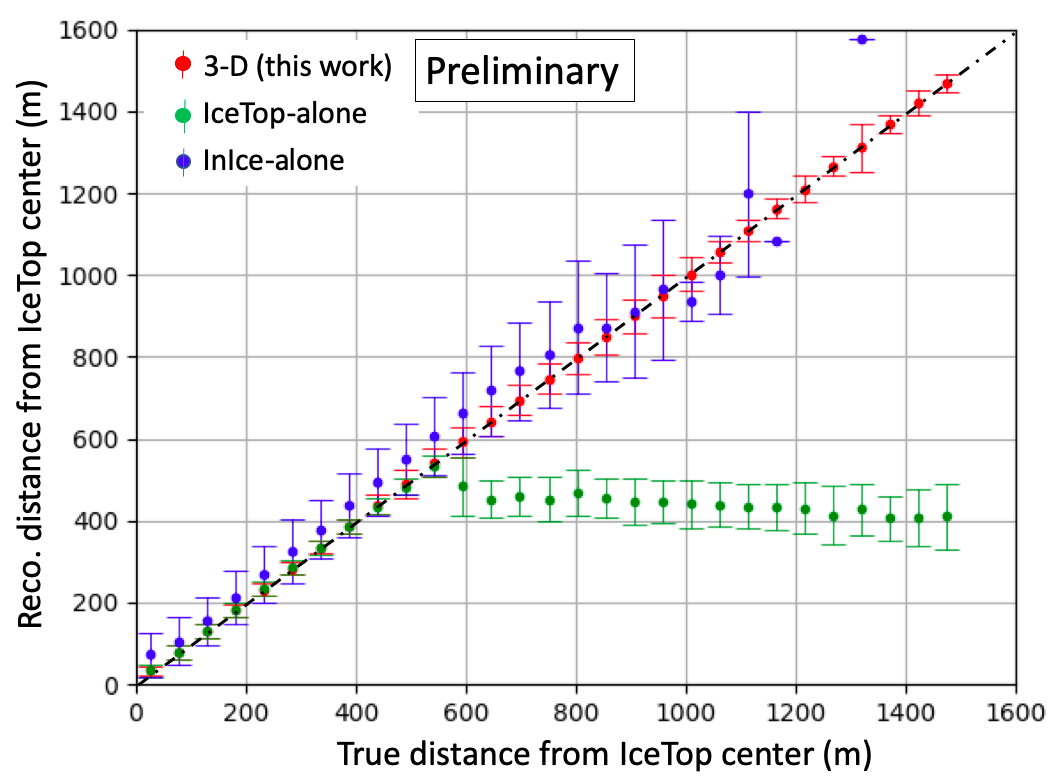}  
\end{center}
\vspace{-1.0pc}
\caption{\label{fig_r} Reconstructed versus Monte Carlo true EAS core distance from the center of 
the IceTop array after QC for all processed proton events. The diagonal black dash dotted lines indicate 
a perfect match between the reconstructed and the true core values. Points above/below the dash 
dotted line correspond to EAS cores dragged artificially away/toward IceTop array center. } 
\end{wrapfigure}
To compare how the MC events can be effectively reconstructed by the IceTop-alone, 
InIce-alone, and the new 3-D reconstruction, the reconstructed EAS core distance from 
the IceTop array center after the QC is compared with the true core distance in Figure~\ref{fig_r}. 
Except several data points with large statistical errors at very high energies due to the small 
number of events, the InIce-only reconstruction has a systematic shift away from the IceTop array 
center. The IceTop-alone reconstruction shifts the reconstructed cores to the inside of the array 
when the real core position is on the edge or outside of IceTop. 
Figure~\ref{fig_r} also shows that the current IceTop QC cannot remove all the events that have 
mis-reconstructed core position. Only the 3-D reconstruction, since it is informed by in-ice 
information as well, does not have this limitation, and is able to correctly find the core of showers 
that land both inside and outside of IceTop. This is an extraordinary improvement that will help 
reduce systematic uncertainties associated with geometric effects. In four primary energy bins 
$10^6 - 10^7$~GeV,  
$10^7 - 10^8$~GeV, 
$10^8 - 10^9$~GeV, and 
$10^9 - 10^{9.5}$~GeV, the successful reconstruction rates 
for IT-Uncontained events by IceTop, InIce, and the 3-D reconstructions after QC are  
(2.4\%, 0.8\%, 0.8\%, 1.1\%)$_{IT}$, 
(63.5\%, 38.1\%, 22.5\%, 13.7\%)$_{II}$,  
(62.6\%, 58.1\%, 43.5\%, 25.3\%)$_{3-D}$. 
By successfully reconstructing a decent portion of IT-Uncontained events, the 3-D reconstruction 
helps reduce statistical uncertainties which is a dominant contributor to the uncertainties at 
high energies. 

\begin{figure}[th]
\begin{center}
\includegraphics[width=0.49\linewidth, angle=0]{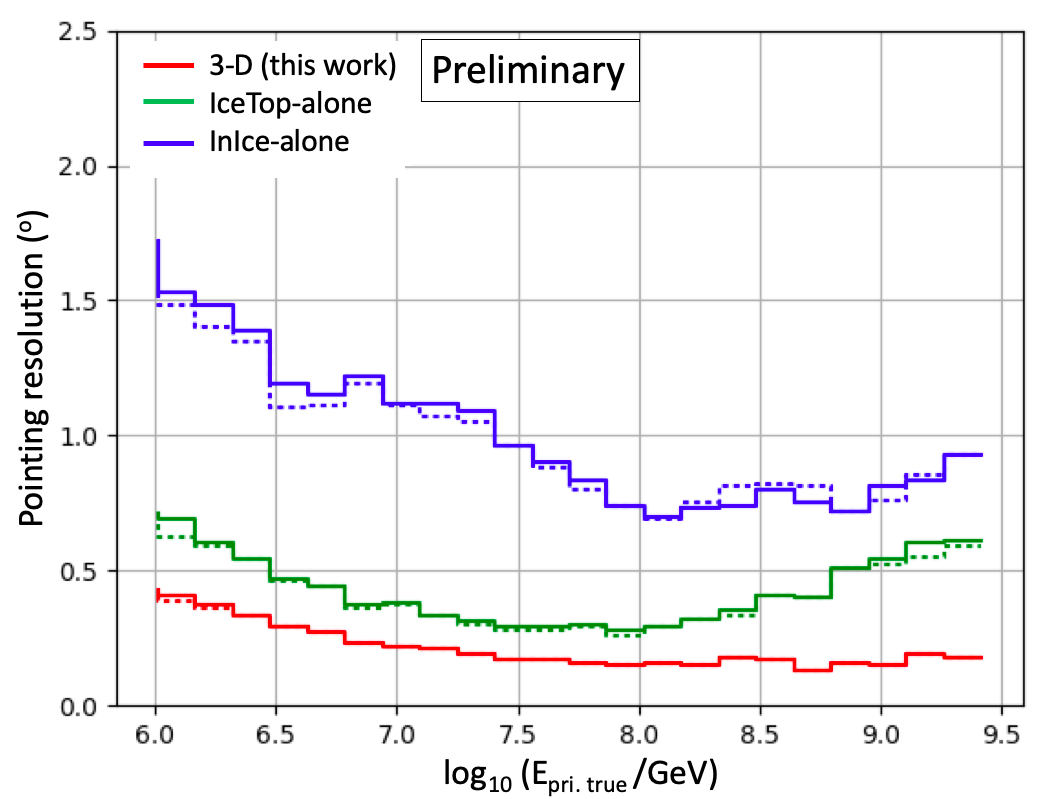} 
\hspace{0pc} 
\includegraphics[width=0.49\linewidth, angle=0]{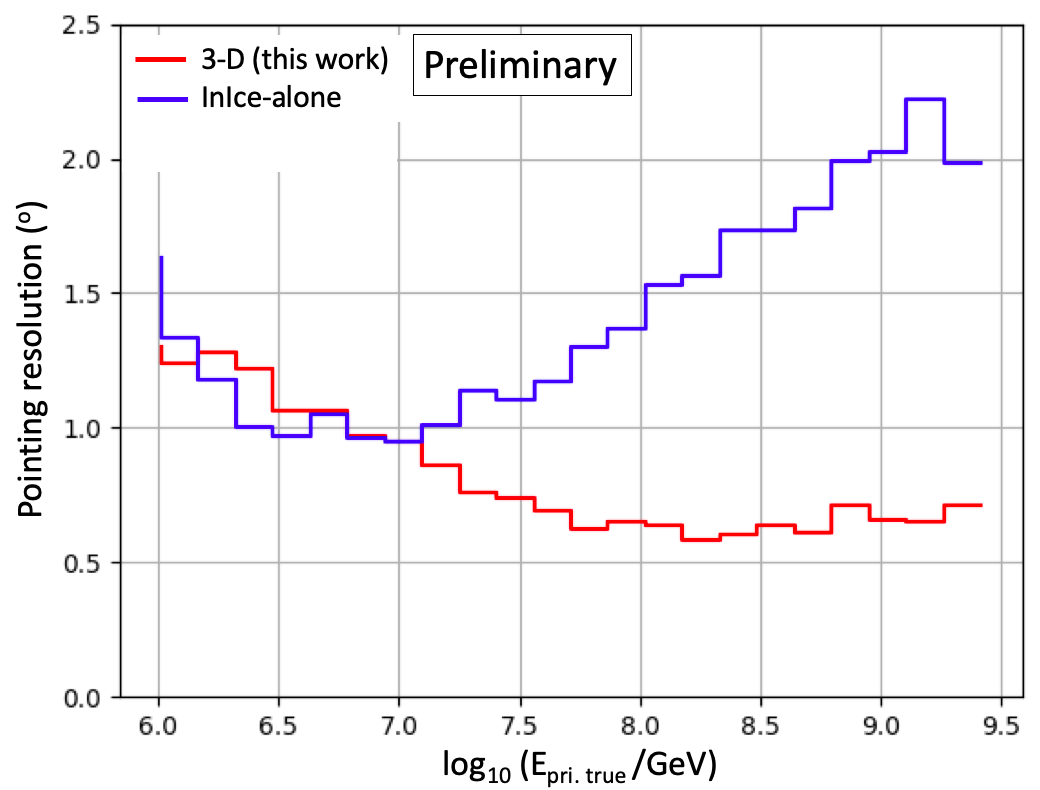} 
\caption{\label{fig_point_resolution}Pointing resolution by three reconstructions for IT-Contained 
(left) and IT-Uncontained (right) proton events. Solid lines are the results from each reconstruction 
after their own QC separately. The dotted lines in the plot on the left represent a comparison in
which all three reconstructions use the same set of events after identical QC.}  
\end{center}
\vspace{-1.0pc}
\end{figure}

The 3-D reconstruction also takes advantage of the long lever arm between the IceTop and the in-ice 
array, which helps reconstruct the CR direction more accurately than either reconstruction can do 
alone. This is particularly true for the IT-Uncontained events. 
Plots in Figure~\ref{fig_point_resolution} show the reconstructed pointing resolution as a function of the 
CR primary energy for events in both groups. 
The value of the resolution is defined 
as the space angle from the true direction which contains 68\% of the reconstructed tracks.
As shown by the red lines in the plot on the left, the 3-D reconstruction does the best for 
IT-Contained events over the entire energy range. The comparison using identical QC for IT-Contained 
events (the dotted lines in the plot of the left) also gives the same result with a slightly improved 
resolution for all three reconstructions. 
For IT-Uncontained events (i.e. plot on the right), the 3-D reconstruction result (red line) is much better than 
the InIce-alone reconstruction except below  $\sim 10^{6.6}$~GeV. This exception is mainly because these 
low energy IT-Uncontained events trigger much fewer IceTop stations. 

\begin{figure}[th]
\vspace{-0.8pc}
\hspace{0.8pc} 
\begin{minipage}{18.0pc}
\begin{center}
\includegraphics[height=12.0pc, angle=0]{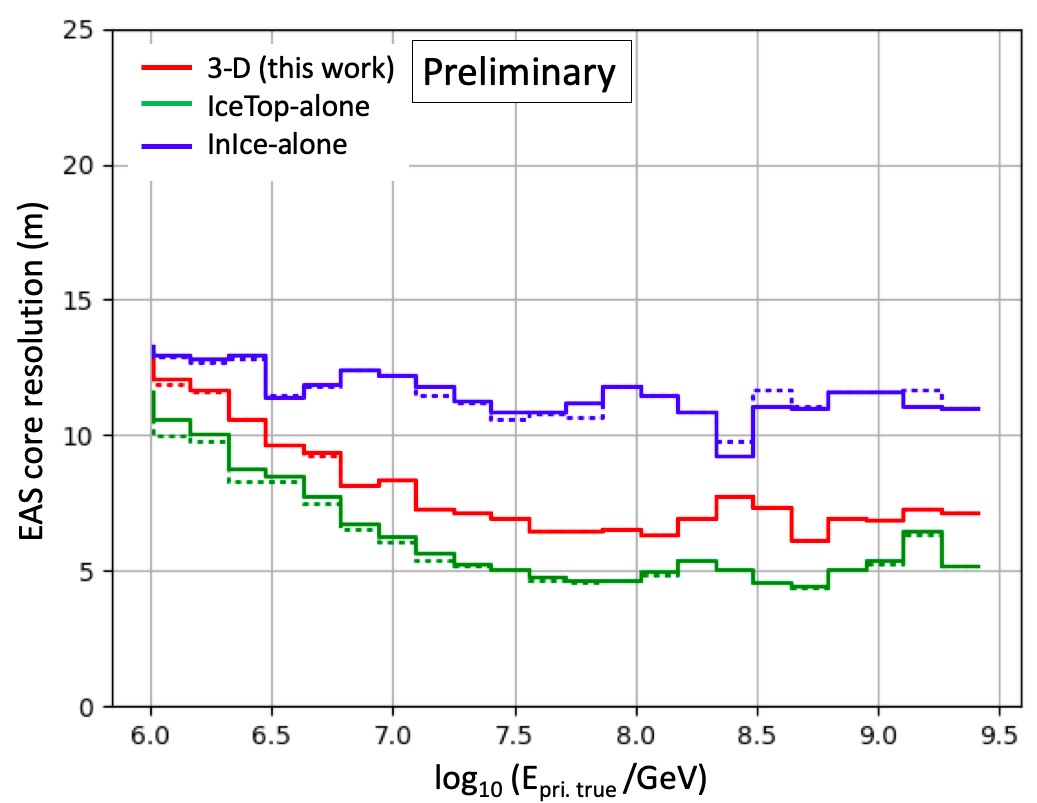} 
\end{center}
\end{minipage}
\hspace{0.7pc} 
\begin{minipage}{13.5pc}
\vspace{-1.5pc}
\caption{\label{fig_core} Comparison of EAS core resolution from IceTop-alone, InIce-alone and the new 3-D 
reconstructions for IT-Contained proton events. Solid lines are the results from each reconstruction after their 
own QC separately. The dotted lines in the plot are the results from these three reconstructions by using the 
same set of events after identical QC.} 
\end{minipage}
\vspace{-0.3pc} 
\end{figure}

Figure~\ref{fig_core} shows the resolution of EAS core location reconstructed separately by these 
three methods for IT-Contained events. The core resolution is defined as the distance from the true 
core which contains 68\% of the reconstructed tracks. 
The core resolution of the 3-D reconstruction is worse than the IceTop-alone reconstruction by 
about 2$\sim$3 meters over the entire energy range being studied, while the InIce-alone reconstruction 
is worse than the IceTop-alone reconstruction by about 3$\sim$8 meters. 
As indicated by the dotted lines in the figure, these differences remain the same in the comparison with 
identical QC for all three reconstructions. 
The deterioration in the 3-D reconstruction core resolution is not a complete surprise because the 
core 
is determined exclusively by the signal sizes in the IceTop detectors.  
Nevertheless, how the deterioration happens while a better angular resolution is achieved 
by the 3-D reconstruction and if there is a way to overcome it still needs more study. 
Since EAS ground particle 
LDF $S(r)$ is very sensitive to the core location (see Eq.~\ref{eq_LDF}), the worse core resolution of the 3-D 
reconstruction also leads to a wider spread in the $S_{125}$-primary energy relation shown below. 

\begin{figure}[th]
\begin{center}
\vspace{0.5pc} 
\includegraphics[width=0.49\linewidth, angle=0]{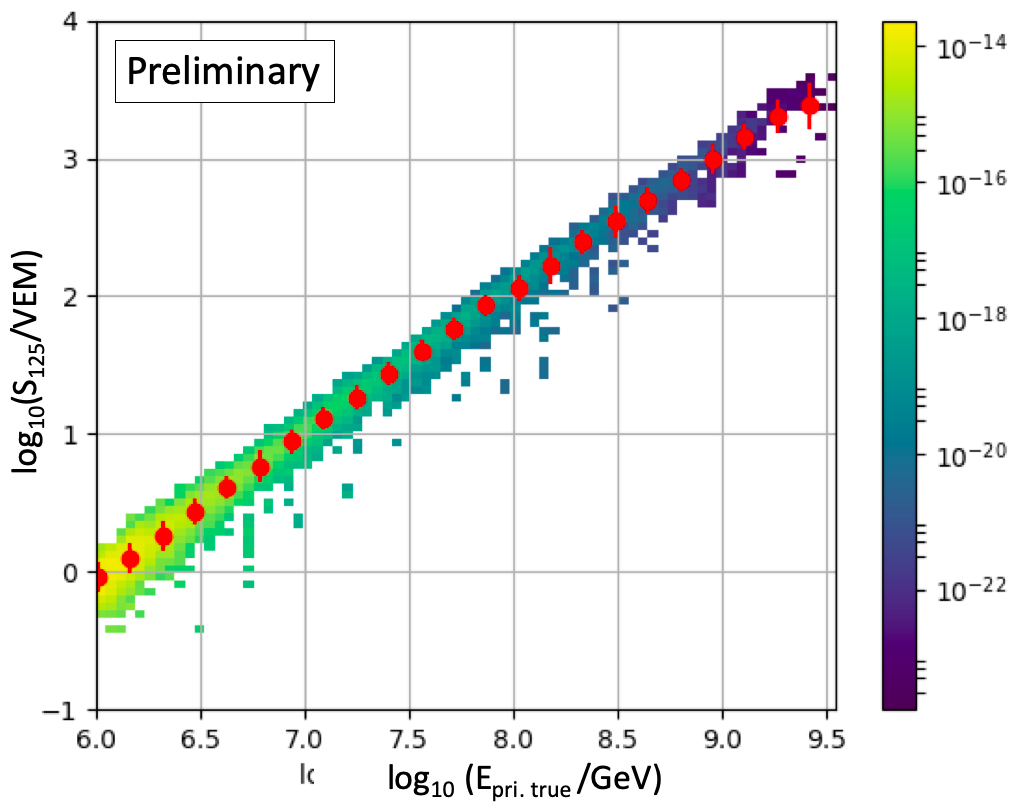} 
\hspace{0pc} 
\includegraphics[width=0.49\linewidth, angle=0]{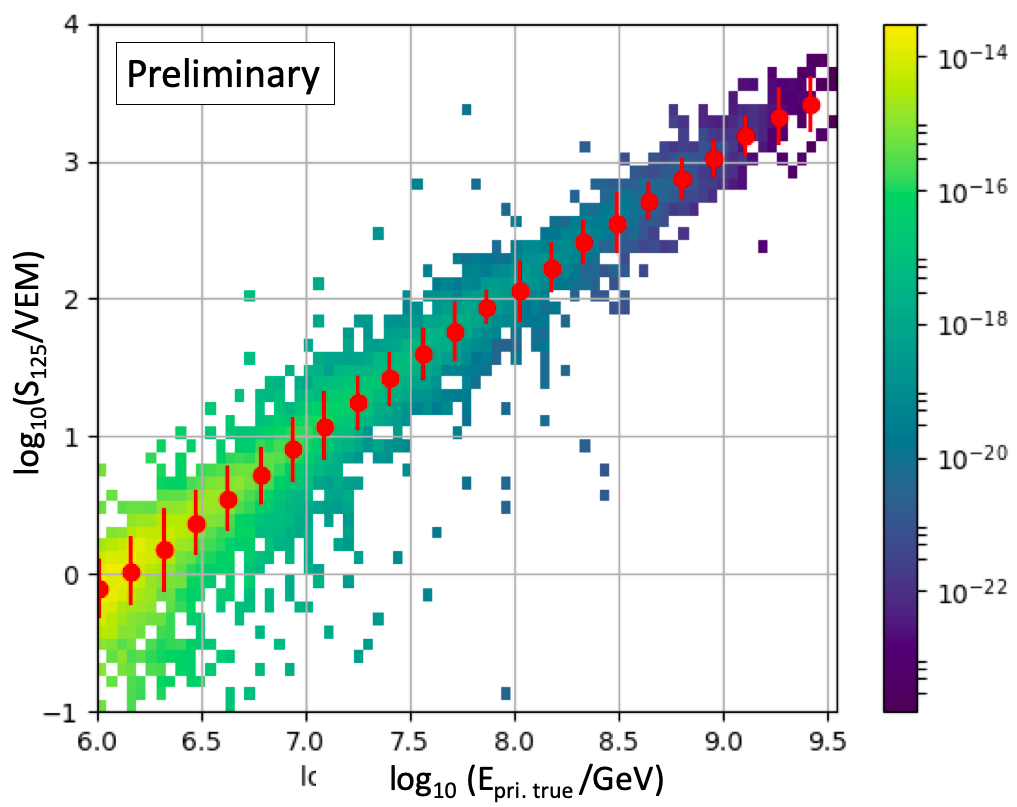} 
\caption{\label{fig_s125_g1} Reconstructed $S_{125}$ versus CR true primary energy for IT-Contained proton 
events after exactly the same QC for both methods. Plots on the left and right are for the IceTop-alone 
reconstruction and the 3-D reconstruction respectively. 
Numbers on the colorbars represent the relative intensity after the spectrum is weighted to $E^{-2.7}$. See 
details in the text.} 
\end{center} 
\vspace{-1.0pc}
\end{figure}

Energy deposition in the surface array by ground particles at 125~m from the EAS core, $S_{125}$, is 
the most important primary energy estimator in IceCube CR analysis. Figure~\ref{fig_s125_g1} shows 
the distribution of the reconstructed $S_{125}$ as a function of CR true primary energy for IT-Contained 
proton events that passed the same QC for both IceTop-alone and the 3-D reconstructions. 
Besides the overall good agreement between these two reconstructions, 
the 3-D reconstruction results in a wider overall spread in the $S_{125}$-energy relationship. 
Since the 3-D reconstruction has a better pointing resolution than IceTop-alone reconstruction for 
IT-Contained events (see plot on the left in Figure~\ref{fig_point_resolution}), the wider spread of 
$S_{125}$ is more likely related to the deteriorated core resolution by the 3-D reconstruction. 
Whether this can be improved by 
fine tuning the likelihood functions, 
optimizing the minimization iteration sequence, 
or by improving QC for the 3-D reconstruction needs more study. 
It is worth to note that, in a "free-style" comparison for which each reconstruction uses its own QC, 
the 3-D reconstruction works better at very high energies by eliminating events misreconstructed by 
IceTop-alone reconstruction. 
Very similar features are also observed for IT-Contained iron events. 

\begin{wrapfigure}{r}{0.5\textwidth}
\begin{center}
\vspace{-1.5pc}
\includegraphics[width=0.50\textwidth]{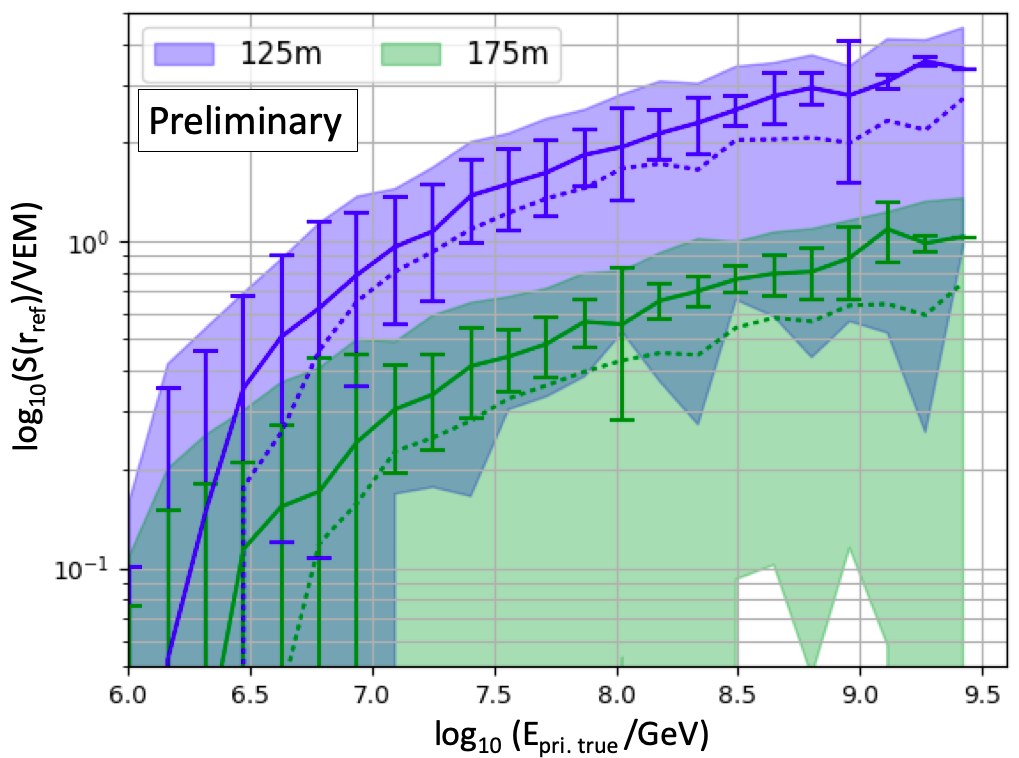} 
\end{center}
\vspace{-0.5pc}
\caption{\label{fig_sr_g2} Value of EAS ground particle LDF at two core distances 
($r_{ref}$=125 m, 175 m) versus CR primary energy for IT-Uncontained proton events. 
Solid lines (with error bars for standard deviation) are for events that have zenith 
angle 0\textdegree<$\theta$<10\textdegree ~and dashed lines (with shaded bands for 
standard deviation) for 20\textdegree<$\theta$<30\textdegree.  
} 
\end{wrapfigure}

Like $S_{125}$ for IT-Contained events, the full reconstruction of IT-Uncontained events also needs 
estimators of primary energy and mass. As a preliminary test of the 3-D reconstruction, 
the $S(r)$ values at different core distances are compared in Figure~\ref{fig_sr_g2} for 
IT-Uncontained proton events in two zenith angle bins. 
The large standard deviation is related to the big zenith bin size and other statistical effects 
(ex. fluctuations in the EAS development, number of events, etc.) 
It can be seen that the 
mean $S(r_{ref})$ is zenith-dependent: events with small zenith angles have a mean $S(r_{ref})$ 
that is systematically higher than more inclined events. 
Our study also showed that the mean $S(r_{ref})$ depends on CR primary mass as well: proton events have 
higher mean $S(r_{ref})$ than iron events at low energies; At high energies, the mean $S(r_{ref})$ of iron 
events measured at core distances larger than 150 m are systematically higher than those of proton events. 
Those observations are qualitatively consistent with the fact that the mean shower maximum 
$X_{max}$ of proton events is larger (i.e. deeper in the atmosphere) by about 100~g/cm$^2$ 
than iron events of the same energy, and vertical proton- and iron-induced EASs that reach 
their $X_{max}$ at IceTop altitude (692~g/cm$^2$) have primary energy of about 10$^8$~GeV 
and 7$\times$10$^9$~GeV respectively. 
For the first time the new 3-D reconstruction enables the study of such relations for IT-Uncontained 
events that cover larger zenith angle than conventional IceTop-InIce coincident analysis which is 
primarily based on IT-Contained events.  
Obviously, more work is needed to quantify these relations in order to find optimal primary energy 
and mass estimators for physics studies with IT-Uncontained events. 

Other investigated quantities include $\beta$ in Eq.~(\ref{eq_LDF}), curvature parameter $c_t$ in 
Eq. (\ref{eq_Curvature}), the average muon bundle energy loss at slant depth 1500 m below the surface, 
and the number of stochastic losses along muon bundle track~\cite{FeuselsPhDThesis}. 
The results of the 3-D reconstruction are either comparable (mostly for IT-Contained events) or 
better (for IT-Uncontained events) than IceTop-alone or InIce-alone reconstructions. 

\vspace{-0.1pc} 
\section{Discussion and outlook}\label{sec_disc} 
\vspace{-0.1pc} 
The new 3-D reconstruction can reconstruct CR events in IceCube with resolution comparable 
with or better than the IceTop-alone or InIce-alone reconstructions in most cases. It also has 
multiple merits. First, the 3-D reconstruction provides additional EAS parameters for events in 
both groups (e.g. curvature parameter $c_t$ in Eq. (\ref{eq_Curvature}), $S_{em}(r)$ and 
$\rho_\mu(r)$, etc.) 
By reconstructing those events landing outside of the IceTop array, i.e. the IT-Uncontained events, 
the 3-D reconstruction not only reduces the statistical uncertainties with significantly more events 
but is also expected to improve the systematics related to EAS development over a broader 
energy range than previous IceTop and in-ice coincident analysis based on IT-Contained events. 
This is because IT-Uncontained events have zenith peaked at about 
25\textdegree ~and extend to about 60\textdegree ~before running out of statistics (versus 
15\textdegree ~and 30\textdegree ~for IT-Contained events), which means that the detector 
can collect data from higher-energy showers that are still near their $X_{max}$, due to the 
increased slant depth through the atmosphere. 

More studies are still needed to optimize the new reconstruction. For example, it would 
be very interesting to see if the additional EAS parameters 
can help reduce uncertainties in the measurement of the mass or energy of CR primary particles; 
Besides the single LDF $S(r)$, the 3-D reconstruction can also fit IceTop data to an 
electromagnetic LDF and a muonic LDF simultaneously (an option provided through 
"Signal Models" in Figure~\ref{fig_FlowChart}), which can be optimized to provide 
additional information for the study of electromagnetic and hadronic processes in EAS 
development. This work has focused on two groups of EAS events that all have high-energy 
muon tracks contained in the in-ice array. To measure the property of high-energy muons that 
are {\it not} contained in the in-ice array is also worth a dedicated study. 
Of course, the quality cuts for the 3-D reconstruction of events in different groups also need 
to be optimized along with these improvements and in the exploration for suitable estimators 
of CR primary mass and energy. 
By realizing the likelihood maximization for observables in both the IceTop and in-ice array 
simultaneously, the 3-D reconstruction provides a necessary tool for these studies to achieve 
a better accuracy for the research with IceCube data. 


\end{document}